\documentclass[reprint,twocolumn,amsmath,showpacs,superscriptaddress,amssymb,aps]{revtex4-1}

\usepackage{graphicx}% Include figure files
\usepackage{graphics}
\usepackage{amsmath}
\usepackage{dcolumn}% Align table columns on decimal point
\usepackage{bm}% bold math

\begin{document}

%\preprint{APS/123-QED}

\title{Strain-driven sign interchange of surface two-dimensional electron and hole gases in KTaO$_3$ thin film}
\author{Xue-Jing Zhang}
\affiliation{Beijing National Laboratory for Condensed Matter Physics, Institute of Physics,
Chinese Academy of Sciences, Beijing 100190, China}
\affiliation{School of Physical Sciences, University of Chinese Academy of Sciences, Beijing 100190, China}
\author{Bang-Gui Liu}\email{bgliu@iphy.ac.cn}
\affiliation{Beijing National Laboratory for Condensed Matter Physics, Institute of Physics,
Chinese Academy of Sciences, Beijing 100190, China}
\affiliation{School of Physical Sciences, University of Chinese Academy of Sciences, Beijing 100190, China}

\date{\today}% It is always \today, today,

\begin{abstract}
Since the discovery of two-dimensional (2D) electron gas in LaAlO3/SrTiO3 interface, 2D carrier gases in perovskite oxides have attracted great attention because they can host many important phenomena and may produce novel functional devices.
Here, we show that there is one pair of surface 2D electron and hole gases in KTaO3 thin film and they can be tuned by applying biaxial stress. For increasing compressive in-plane strain, the 2D carrier concentrations decrease down to zero, and then a new pair of surface 2D electron and hole gases are formed and the carrier signs are interchanged.
Our analysis indicates that this carrier sign interchange happens because the increasing compressive strain reverses the slope of monolayer-resolved electrostatic potential along the [001] direction. Furthermore, we also present strain-dependent carrier concentrations and effective masses and  their thickness dependence, and show that the surface 2D carrier gases and their strain-driven sign interchange can persist even in the presence of overlayers and epitaxial substrates. These phenomena should be useful to design novel functional devices.
\end{abstract}

\maketitle

%\tableofcontents

\section{INTRODUCTION}

The discovery of a two-dimensional electron gas (2DEG) at the interface between two oxide insulators,
SrTiO$_3$ (STO) and LaAlO$_3$ (LAO), has attracted significant attention\cite{1}. Polar discontinuity between
polar and nonpolar insulators accounts for the driving force for the accumulation of free charge at these
interfaces\cite{2,add9}. This has also triggered great interest in designing 2DEG at oxide interfaces composed
of perovskite ferroelectrics, using the ferroelectric polarization as the source for polar discontinuity,
such as PbTiO$_3$(PTO)/STO\cite{3}, BaTiO$_3$/STO\cite{4}, BaTiO$_3$/PTO\cite{5}, KNbO$_3$/ATiO$_3$
(A= Sr. Ba, Pb)\cite{6}, etc. Paraelectric-to-ferroelectric transition in STO films can be triggered by an epitaxial strain\cite{11,12}. However, the surface electronic states observed at the vacuum-cleaved
surface of STO open another avenue for the understanding and fabrication of 2DEG in transition-metal oxides\cite{7,8}.
It was argued that this surface electronic states arises from surface oxygen vacancies\cite{9}. Very recently, it was shown that a pair of 2D carrier (both electron and hole) gases can be realized in STO/LAO/STO trilayers\cite{add8}

Tantalates and niobates are also prototypical examples of polar perovskites, and many of them exhibit ferroelectricity\cite{f1}.
Recent angle-resolved photoemission spectroscopy (ARPES) experiments
reported the realization of a 2DEG directly at the vacuum-cleaved surface of KTaO$_3$ (KTO)\cite{9,10}. Such surface electronic structures can be modified by Ar$^+$ bombardment\cite{add7}. KTO (001) thin film can be considered to consist of alternately negatively-charged (KO)$^-$ monolayers and positively-charged  (TaO$_2$)$^+$ monolayers along the [001] direction. Two-dimensional electronic structures can be formed in KTO-based interfaces, such as KTO/GdTiO$_3$ and LAO/KTO interfaces\cite{add6,add1}.
The existence of strain-induced ferroelectric order in quantum paraelectric KTO has been
demonstrated by experimental investigations of epitaxial KTO films\cite{13}. The two-dimensional electron (hole) gas can be naturally formed in high-quality TaO$_2$-terminated (KO-terminated) surface\cite{s1}.

Strain (stress) is a wonderful approach to manipulate crystal structures of perovskite oxides and thus control their electronic structures and functional properties\cite{add9,11,12,13,add3}. Here, through first-principles
calculations, we show that biaxial stress can be used to create tunable surface 2DEG and 2DHG and then drive sign interchange of the 2D carrier gases in KTO thin film. The microscopic mechanism is elucidated, and key parameters and thickness dependence are presented. Effects of overlayers and epitaxial substrates are also explored. More detailed results will be presented in the following.

\begin{figure*}[!tbp]
\centering  % Requires \usepackage{graphicx}
\includegraphics[clip, width=16cm]{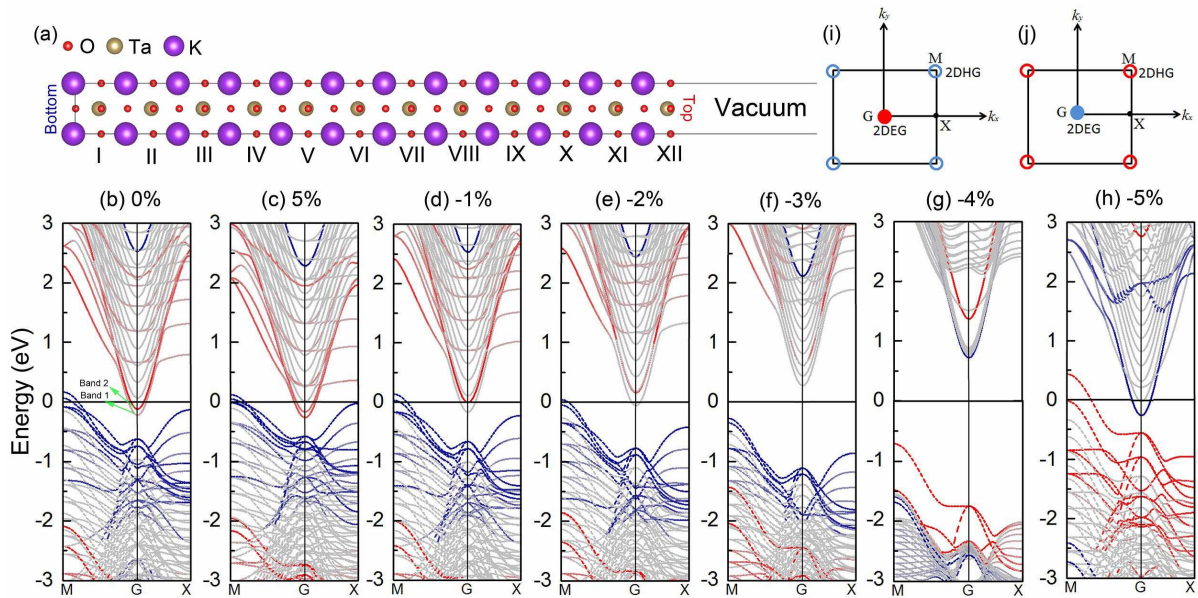}
\caption{(Color online) (a) Side view of the optimized atomic structure of (KTO)$_{12}$ slab at $\varepsilon_s$=0\%.
(b)-(h) Band structures of the (KTO)$_{12}$ slab at different strain values: $\varepsilon_s$=0\%, 5\%, -1\%, -2\%,
-3\%, -4\%, and -5\%. The red and blue lines in the conduction bands describe the energy bands originating from Ta
atoms of the monolayers labelled with 'XII' and 'I', respectively. The red and blue lines in the valence bands
indicate the energy bands originating from O atoms in the top and bottom surfaces, respectively. (i)-(j) The
corresponding 2DEG (filled circles) and 2DHG (hollow circles) regions in the Brillouin zone for $\varepsilon_s$=0\%
and -5\%, respectively, and red and blue correspond to the top and bottom surfaces.}\label{edge}
\end{figure*}

\section{COMPUTATIONAL DETAILS}

Our first-principles calculations are performed using the projector-augmented wave method within the
density-functional theory\cite{30,31}, as implemented in the Vienna Ab-initio Simulation Package (VASP)\cite{32,33}.
We use the generalized gradient approximation for solid  (PBEsol)
through revising the Perdew-Burke-Ernzerhof (PBE) functionals for the exchange-correlation functionals, because it usually produces better results than the usual PBE  for solids and their surfaces\cite{34,35}. The KTO structures
are fully optimized  using a $\Gamma$-centered $4\times 4\times 1$ k-grid, with in-plane lattice constant constrained to the experimental lattice constants. The electronic structure calculations were performed by using
a $\Gamma$-centered $8\times 8\times 1$ k-grid. The plane wave energy cutoff is set to 500 eV. Our convergence standard requires that the Hellmann-Feynmann force
on each atom is less than 0.01 eV/\AA{} and the absolute total energy difference between two successive loops is
smaller than $10^{-5}$ eV. A 20 \AA{} thick vacuum layer ensures removing possible artificial results in the slab geometry. The strain is defined
as $\varepsilon_s = (a-a_0)/a_0\times100\%$, where $a_0$ is the experimental lattice constant of KTO and $a$ is
the in-plane lattice constant under strain. Given a strain value, the out-of-plane lattice constant and all the
internal atomic positions are allowed to relax sufficiently in optimization.

\section{RESULTS AND DISCUSSION}

\subsection{Surface 2D carrier gases in KTO thin film}

The KTO bulk has experimental lattice constant $a_e=3.989$ \AA. The optimized lattice constants of KTO with PBE and PBEsol
are 4.034 and 3.996 \AA, being larger than the experimental lattice constant by 1.13\% and 0.18\%, respectively,
and therefore we shall use PBEsol in the following. We build up a KTO slab model to
describe a KTO thin film. The slab consists of $m$=12 KTO unit cells in the vertical [001] direction. Fig. 1(a) shows the
optimized structure of the KTO slab at $\varepsilon_s$=0\%. The energy bands of the equilibrium KTO
layer ($\varepsilon_s$=0\%) are presented in Fig. 1(b). The red and blue solid lines in the conduction bands indicate
the bands originating from Ta atoms in the monolayers labelled with 'XII' and 'I', respectively. It is clear that the
conduction bands are filled by some electrons, which realizes a two-dimensional electron gas (2DEG) in the top surface.
The red and blue dash lines in the valence bands indicate the bands originating from O atoms in the top and bottom
surfaces of two unit cells in the KTO layer, respectively. The valence bands are not completely filled. Accordingly,
there exists a two-dimensional hole gas (2DHG) originating from the bottom surface. These are visualized in the
Brillouin zone in Fig. 1(i).
The monolayer-resolved DOS curves for Ta $d$ and O $p$ in the KTO layer at $\varepsilon_s$=0\% are presented in Fig. 2(a).
They can be used to confirm that the 2DEG and 2DHG are in the top and bottom surfaces, respectively.
The pair of the 2D carrier gases in the two surfaces reflects the fact that
the KTO layer has the polar property, (KO)$^-$...(TaO$_2$)$^+$, along the [001] direction.

Orbital-resolved DOS analysis shows that the 2DEG is made mainly from surface Ta $d_{xy}$ states at the top end of the positively-charged (TaO$_2$)$^+$) monolayers, because they
are farther from the O ligands than the other $d$ orbitals near the surfaces, experiencing less repulsion. The 2DHG
is made mainly from the three surface O $p$ states at the bottom end of the negatively-charged (KO)$^-$ monolayer, where Ta bonds with neighboring O atoms in three different directions, making two of the three orbitals easier to make holes. This pair of surface 2D carrier gases is consistent with that in the LAO layer\cite{add8}.

\begin{figure}[!tbp]
\centering  % Requires \usepackage{graphicx}
\includegraphics[clip, width=8.6cm]{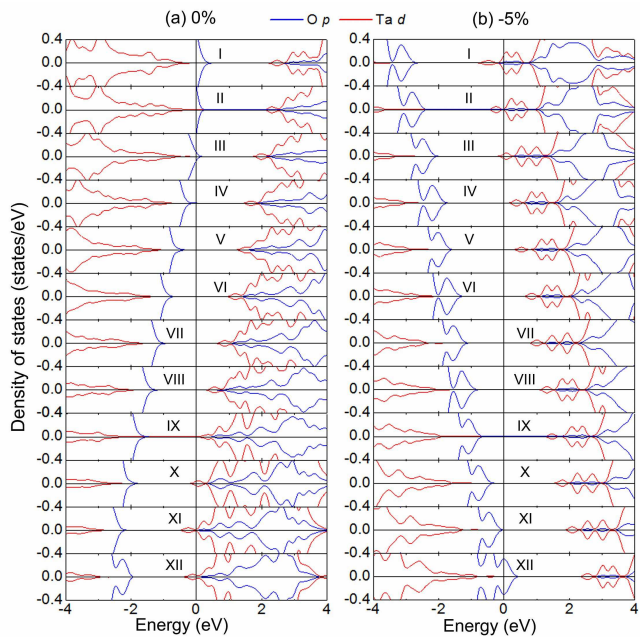}
\caption{(Color online) The DOS curves of the Ta $d$ and O $p$ of the unit-cell layers labelled from 'I' to 'XII'
of (KTO)$_{12}$  unit cell for $\varepsilon_s$=0\% (a) and -5\% (b). }\label{edge}
\end{figure}

\subsection{Strain-driven sign interchange of the surface carriers of electrons and holes}

When a biaxial compressive stress is applied to the KTO layer, there will be a biaxial compressive strain in the
plane perpendicular to the [001] axis and a tensile [001] strain determined by zero stress condition for the axis.
For convenience, we use the in-plane strain to parameterize the strained system, describing the stress and the [001] strain\cite{add2,add4}, in the following. When
the in-plane strain is tensile (larger than zero), there are no substantial changes in the energy bands, as shown
for $\varepsilon_s$=5\% in Fig. 1(c). In contrast, large changes are observed in the cases of the compressive strains. We present in Figs. 1(d)-(h) the calculated electronic band structures of the
KTO layer with the in-plane strain $\varepsilon_s$ values of -1\%, -2\%, -3\%, -4\%, and -5\%. It is clear that
with increasing the compressive strain (up to $\varepsilon_s$=-4\%), the conduction band bottom moves upward and the
valence band top downward. These make the KTO layer lose the surface 2D carrier gases and become insulating
when $|\varepsilon_s|$ is larger than 2\% (up to 4\%). Surprisingly, when the compressive strain increases further
from -4\%, another pair of surface 2D carrier gases can be formed. We plot the bands for $\varepsilon_s$=-5\% in Fig. 1(h),
which clearly shows that there are a pair of surface 2D carrier gases.

Our spectral weight analysis shows that for $\varepsilon_s=-5$\%, however, the 2DEG can be attributed
to the bottom surface and the 2DHG to the top surface, as shown by the red and blue lines in Fig. 1(h).
We make these more visible in Fig. 1(j). We also present the monolayer-resolved DOS curves for Ta $d$ and O $p$ in
the KTO layer for $\varepsilon_s=-5$\% Fig. 2(b). With all these, it can be clearly seen that the carrier sign of
the 2DEG and 2DHG at the two surfaces are interchanged when the in-plane strain is switched from 0\% to -5\%.

\begin{figure*}[!htbp]
\centering  % Requires \usepackage{graphicx}
\includegraphics[clip, width=16cm]{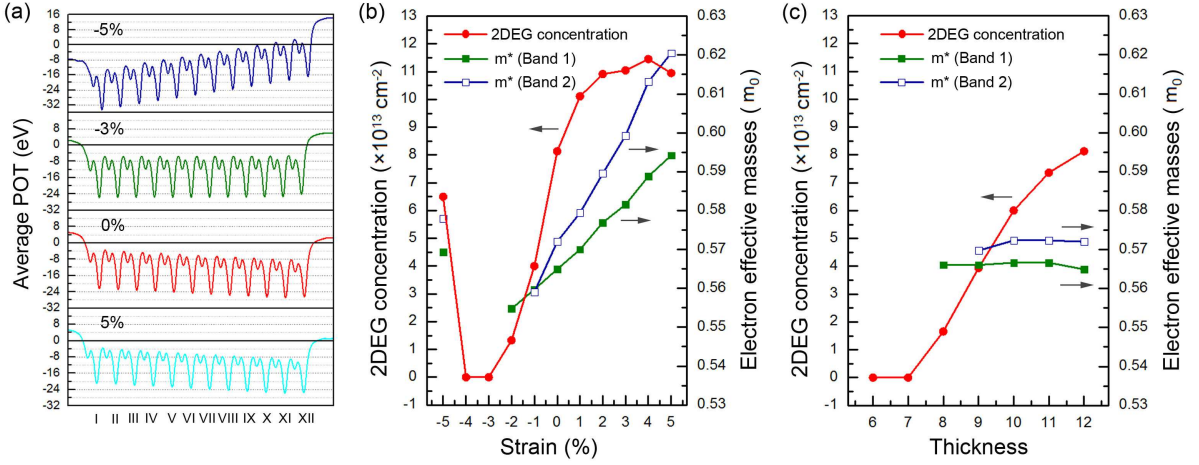}
\caption{(Color online) (a) The planar averaged electrostatic potential across the (KTO)$_{12}$ slab for
$\varepsilon_s$=-5\%, -3\%, 0\%, and 5\%. (b) 2DEG concentration and electron effective masses of (KTO)$_{12}$ as
functions of strain $\varepsilon_s$. (c) 2DEG concentration and electron effective masses of (KTO)$_{n}$ ($n=6\sim 12$)
as functions of the layer thickness ($n$, in unit cell). }\label{edge}
\end{figure*}

\subsection{Monolayer-resolved electrostatic potential}

In order to elucidate the evolution trend of the electronic structure, we present the monolayer-resolved averaged electrostatic
potential across the KTO layer for $\varepsilon_s$=-5\%, -3\%, 0\% and 5\% in Fig. 3(c). It is clear that there
exists a descendent slope of electrostatic potential from the left to the right at $\varepsilon_s$=0\%. This can
be attributed to the polarity, (KO)$^-$...(TaO$_2$)$^+$, of the KTO layer along the [001] direction. This slope of
electrostatic potential remains almost unchanged when $\varepsilon_s$ increases, up to 5\%. When we make the strain
compressive (up to -2\%), the slope of electrostatic potential smoothly becomes smaller (very small for $\varepsilon_s$=-2\%).
When the compressive strain reaches $\varepsilon_s$=-3\%, the slope is almost equivalent to zero, and therefore the
electrostatic potential becomes almost flat and the KTO layer is insulating.
When we further increase the compressive strain ($\varepsilon_s <$-3\%), the slope of electrostatic potential is reversed
and its absolute value increases with the compressive strain. We have a substantially increscent slope of electrostatic
potential from the left to the right at $\varepsilon_s$=-5\%, which reflects the increasing cation-anion polarization,
as shown in Fig. S1.

It can be seen that, when the in-plane strain $\varepsilon_s$ changes from 5\% to -5\%, the relative displacement
along z axis of cation M (M: K, Ta) with respect to the anion O in each unit cell of the KTO layer persists to increase.
For the in-plane strain $\varepsilon_s$=-3\%, with the out-of-plane lattice constant 3.869 \AA, the cation-anion
polarization can balance with the polarity in the KTO layer along the [001] direction, and these diminish the 2D
carrier gases and make the systems become insulating.

\subsection{Key parameters and thickness dependence}

2DEG concentrations of up to $3\times$10$^{13}$ cm$^{-2}$ can be achieved in STO/LAO system, and 3$\times$10$^{14}$
cm$^{-2}$ have been demonstrated at STO/GdTiO$_3$ interfaces\cite{36,37}. For the KTO layers, the calculated 2DEG
concentrations $n_e$ and electron effective masses ($m_{e1}$, $m_{e2}$) as functions of $\varepsilon_s$ are shown in
Fig. 3(b). For $\varepsilon_s$=0\%, we have $n_e=0.8\times$10$^{14}$ cm$^{-2}$ and $m_{e1}=0.565m_0$ ($m_0$ is the mass
of free-electron). The tensile strain can increase the 2DEG concentration and enlarge the electron effective mass.
With the tensile strain $\varepsilon_s$=5\%, they are  $n_e=1.1\times$10$^{14}$ cm$^{-2}$ and $m_{e1}=0.594m_0$. $m_{e2}$
is larger than $m_{e1}$. With the compressive strain, both the 2DEG concentration and the electron effective masses
become smaller. Especially, for $\varepsilon_s=-3\sim -4$\%, the system is an insulator with band gap of $0.53\sim 1.4$ eV.
For $\varepsilon_s$=-5\%, the system becomes conductive, and its surface 2D carrier gases take reversed signs, as shown
in Fig. 2(b).

We also study thickness-dependent properties in the KTO layer. For (KTO)$_6$ with strain $\varepsilon_s$ changing
from -5\% to 5\%, calculated electronic band structures are presented in Fig. S2. In contrast to the (KTO)$_{12}$,
the (KTO)$_6$ layer is an insulator for $\varepsilon_s \le 2$\%. Furthermore, we investigate the variation of 2DEG
concentrations and electron effective masses in (KTO)$_n$ ($n=6\sim 12$) for $\varepsilon_s$=0\%, showing the
thickness-dependent results in Fig. 3(c). It can be seen that the 2DEG concentrations substantially decreases with the
thickness, but the electron effective masses have small thickness dependence.  For the small thickness $n\le 7$,
the system is insulating and there are no carriers. Therefore, $n_c=8$ is the critical thickness of the KTO layer
with $\varepsilon_s$=0\% for the surface 2D carrier gases and interesting phenomena concerned.

\begin{figure}[!tbp]

\centering  % Requires \usepackage{graphicx}
\includegraphics[clip, width=8.6cm]{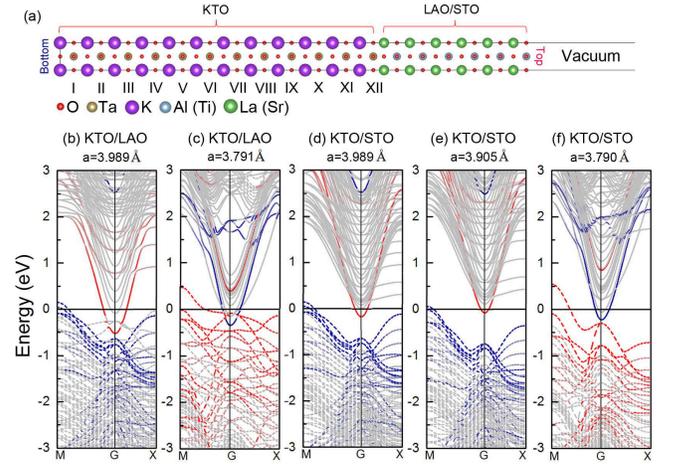}
\caption{(Color online) (a) Side view of crystal structure of (KTO)$_{12}$/(LAO)$_6$ or (KTO)$_{12}$/(STO)$_6$.
(b)-(c) The band structures of (KTO)$_{12}$/(LAO)$_6$ for the in-plane lattice constants: $a_{\rm KTO}=3.989$ \AA{}
and  $a_{\rm LAO}=3.791$ \AA{}. (d)-(f) The band structures of (KTO)$_{12}$/(STO)$_6$ for $a_{\rm KTO}=3.989$ \AA{},
$a_{\rm STO}=3.905$ \AA{}, and $a_{\parallel}=3.790$ \AA{}, respectively.}\label{edge}
\end{figure}

\subsection{KTO based heterostructures}

Now, we discuss KTO-based heterostructures for potential applications. We consider KTO/LAO and KTO/STO bilayers
with TaO$_2$/LaO and TaO$_2$/SrO interfaces, respectively. The experimental lattice constants of bulk KTO, LAO,
and STO are $a_{\rm KTO}=3.989$ \AA{}, $a_{\rm STO}=3.905$ \AA{}, and  $a_{\rm LAO}=3.791$ \AA{}, and corresponding
calculated band structures are presented in Fig. S3. Our computational models are (KTO)$_m$/(LAO)$_n$
and (KTO)$_m$/(STO)$_n$ with $m=12$ and $n=6$, plus a vacuum layer of 20\AA{}, as shown in Fig. 4(a). For an in-plane
lattice constant $a_{\parallel}$, we optimize the out-of-plane lattice constant $a_{\perp}$ and internal atomic
positions. The band structures of (KTO)$_{12}$/(LAO)$_6$ are calculated for $a_{\parallel}=a_{\rm KTO}$
and $a_{\rm LAO}$, and presented in Fig. 4(b)-(c). Here, the (LAO)$_6$ layer is too thin to host 2D carrier gases
for $\varepsilon_s=5$\%.
Although there exists the LAO overlayer, the sign reversal of the 2D carrier gases in the KTO layer is observed
when $a_{\parallel}$ is switched from $a_{\rm KTO}$ to $a_{\rm LAO}$ ($\varepsilon_s$=-5\%).
For $a_{\parallel}=a_{\rm KTO}$, this model can be
used to describe a thin KTO film covered with an ultra-thin LAO overlayer\cite{add1}.
For (KTO)$_{12}$/(STO)$_6$ system, the band structures for $a_{\parallel}=a_{\rm KTO}$ and $a_{\rm STO}$ are presented
in Fig. 4(d)-(e). For this $a_{\parallel}$ value, there are no 2D carrier gases in the STO layer,
and the 2D carrier gases in the KTO layer are almost the same as those in the pure KTO layer. Similar to the case
of KTO/LAO, this KTO/STO model with $a_{\parallel}=a_{\rm KTO}$ can simulate a  KTO thin film covered
with a ultra-thin epitaxial STO overlayer.
Furthermore, for $a_{\parallel}=a_{\rm STO}$ this KTO/STO model  can be used to describe an epitaxial
thin KTO film on STO substrate because the
critical thickness for 2D carrier gases in the STO layer is larger than 6 and there are no surface states at the
STO surface. When $a_{\parallel}$ further decreases
to $3.790$\AA{}, the in-plane strain in the KTO layer reaches -5.0\%, and the sign reversal of the 2D carrier gases
in the KTO layer can be clearly seen, as shown in Fig. 4(f).
Therefore, the strain-driven sign reversal of the surface 2D carrier
gases in the KTO layer remains true even in the presence of ultra-thin epitaxial overlayers or epitaxial substrates.

%\subsection{Further insight}

\section{CONCLUSION}

In summary, we have systematically investigated the structural and electronic properties of KTO thin film in the presence of
strain. We observe one pair of surface 2D electron and hole gases in the KTO thin film. The electron gas is at the top surface and the hole gas at the bottom surface. These 2D carrier gases are tunable through applying biaxial stress. For tensile in-plane strain, the carrier concentrations increase with the strain. When we apply the increasing compressive strain, the 2D carrier concentrations decrease down to zero, there is a metal-semiconductor transition, and then happens a semiconductor-metal transition and a new pair of surface 2D electron and hole gases are formed, but the carriers experience a strain-driven sign interchange, with the electron gas switched to the bottom surface and the hole gas to the top surface.
Our analysis indicates that this transition happens because the increasing compressive strain reverses the slope of monolayer-resolved electrostatic potential along the [001] direction. We also present strain-dependent carrier concentrations and effective masses, and explore their thickness dependence. Further study shows that the surface 2D carrier gases and their strain-driven sign interchange can persist even in the presence of overlayers and epitaxial substrates. These novel phenomena may open a door to novel functional devices.

\section*{ACKNOWLEDGMENTS}

This work is supported by the Nature Science Foundation of China (Grant No. 11574366), by the Department of Science
and Technology of China (Grant No. 2016YFA0300701), and by the Strategic Priority Research Program of the Chinese
Academy of Sciences (Grant No. XDB07000000). The calculations were performed in the Milky Way \#2 supercomputer system
at the National Supercomputer Center of Guangzhou.

\end{document}